\documentclass[final,1p,times]{elsarticle} 

\usepackage{graphicx}
\usepackage{amssymb} 
\usepackage{amsthm} 
\usepackage{lineno}

\def\pe{${p^{e}_{T}}$}                                                                                 
\def\pt{${p_{T}}$}                                                                                 
\def\snn{$\sqrt{s_{_{NN}}}$ =}                                                                      
\def\p{$p\mathord{+}p$}                                                                                  
                                                                                       
\def\Au{Au$\mathord{+}$Au}

\def\Raa{$R_{AA}$}                                                                                       
\def\Rb{$R^{b \rightarrow e}_{AA}$}                                                                                       
\def\Rc{$R^{c \rightarrow e}_{AA}$}                                                                                       
\def\K{$K_{e3}$}                                                                                       
                                                                                       
\def\Fb{$F^{b}$}

\journal{Nuclear Physics A} 
\begin{document}
\begin{frontmatter} 
\title{Probing Hot and Dense Matter with Charm and Bottom Measurements with PHENIX VTX Tracker}   
\author{Rachid Nouicer (for the PHENIX\fnref{col1} Collaboration)}
\fntext[col1] {A list of members of the PHENIX Collaboration and acknowledgments can be found at the end of this issue.}
\address{Physics Department, Brookhaven National Laboratory, Upton, NY 11973, USA.}

\begin{abstract}
We present the first measurements of the nuclear modification factor
($R_{AA}$) for flavor-separated $b$, $c$-quark electrons in
\Au\ collisions at \snn\ 200 GeV. The newly installed Silicon Vertex
Tracker is used to measure the distance of closest approach
distributions of electrons at midrapidity ($\mid$$\eta$$\mid$$
<$~0.35) over the transverse momentum range 1 $<$\pe$<$ 5
GeV/$c$. From this, the relative fraction of bottom ($b$) and charm
($c$) quarks is determined in both the \p\ and \Au\ collision systems,
which form the basis of the measured $R_{AA}$. In \p, we observe that
a FONLL perturbative QCD calculation of
$b\rightarrow$$e$/($c\rightarrow$$e$+$b\rightarrow$$e$) ratio is in
good agreement with the data. In \Au, the data imply a large
suppression of $b$$\rightarrow$$e$ or a large modification of $B$
meson \pt\ distributions, which implies very interesting physics of
$B$ mesons in \Au\ collisions.
\end{abstract} 

\end{frontmatter} 
\linenumbers
\section{Physics Motivation}
In relativistic heavy ion collisions at RHIC, heavy quarks (charm,
$c$, or bottom, $b$) are expected to be primarily created from initial
hard parton scatterings~\cite{Lin95} and carry information from the
system at an early stage. The interaction between heavy quarks and the
medium is sensitive to the medium dynamics, therefore heavy quarks are
suggested as an ideal probe to quantify the properties of the strongly
interacting QCD matter. Heavy quark production has been studied by the
PHENIX experiment via the measurement of electrons from semi-leptonic
decays of hadrons carrying charm (noted $c$) or bottom (noted $b$)
quarks. A large suppression and strong elliptic flow of single
electron heavy flavor has been observed in Au+Au collisions at
\snn\ 200\,GeV~\cite{PHSUP}.  This suppression is found to be similar to
that of light mesons which implies a substantial energy loss of fast
heavy quarks while traversing the medium. The strong flow implies that
the same heavy quarks are in fact sensitive to the pressure gradients
driving hydrodynamic flow -- giving new insight into the strongly
coupled nature of the QGP fluid at these temperatures.  For these
earlier results, PHENIX was not able to distinguish electrons from $c$
and $b$ quarks. In order to understand medium effects in more detail,
it became imperative to directly measure the nuclear modification, and
the flow, of $c$ and $b$ separately. Based on this motivation, in
December 2010, the PHENIX Collaboration opened a new era for measuring
heavy flavor at RHIC by installing a new detector called the Silicon
Vertex Tracker (VTX).

The new VTX detector benefits three areas for PHENIX heavy flavor
measurements.  First, by selecting electrons with a distance of
closest approach (DCA) to the primary vertex larger than
$\sim$100\,$\mu$m, the photonic electron background is suppressed by
orders of magnitude.  This suppression results in a clean and robust
measurement of heavy flavor production in the single electron
channel. Secondly, as the lifetime of mesons containing  bottom is
significantly longer than those containing charm, the detailed DCA
distribution from the VTX allows to disentangle charm from bottom
production over a broad \pt\ range. Thirdly, a DCA cut to remove
hadrons reduces the combinatorial background of $K\pi$ such that a
direct measurement of $D$ mesons through this decay channel will be
possible. In this paper, we present the first PHENIX measurements
using the VTX to measure the DCA of electrons at midrapidity.  From
this, the yield ratio of single electrons from bottom to those from
all heavy flavor, as well as the nuclear modification factor of charm
and bottom separately, are presented using minimum-bias (MB) \Au\ and
\p\ collisions at \snn~200\,GeV obtained in the 2011 and 2012 RHIC
runs.
\begin{figure}[!]                                      
\includegraphics[width=1\textwidth]{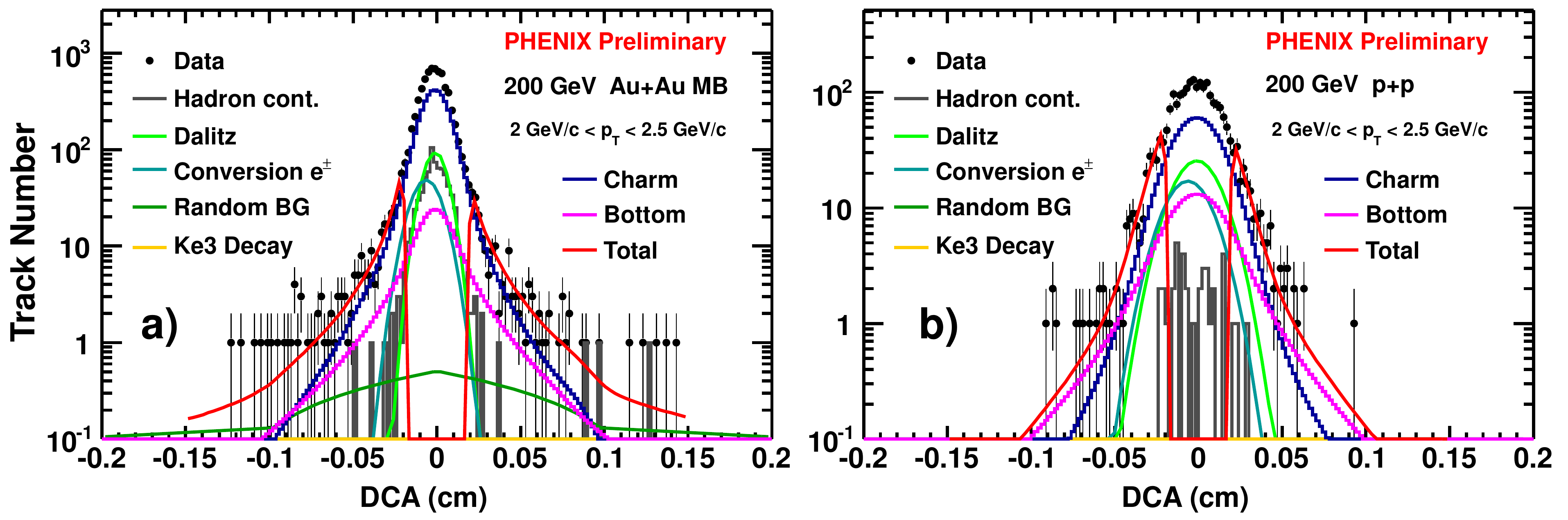}
\vskip -0.15cm
\caption{\small \label{fig1} DCA distribution of
  2$<$\pe$<$2.5\,GeV/$c$ electrons measured at \snn\,200\,GeV in (a)
  minimum-bias \Au\ and (b) \p\ collisions.  The curves represent the
  decomposed contribution of charm, bottom, photonic background, and
  other components. The total curve is shown for the fit range used to
  extract the b and c ratio (see text for details).}
\end{figure}                                      
\section{PHENIX Central Detectors}
Electron identification in PHENIX utilizes the two central-arm
detectors~\cite{PHE03}. Until 2010, a combination of tracker and
electromagnetic calorimeter systems allowed for electron, photon, and
hadron measurements over the range $\mid$$\eta$$\mid$$<$0.35 and
$\Delta\phi$\,=\,$\pi$/2 in azimuth.  In 2010, the central detector
was upgraded to include the VTX ready for Run 2011.  Starting from the
beam line, the complete detector now comprises the new VTX, which is
followed by another tracker system made of two sets of drift chambers
and a pad chamber (PC1). Outside of the PC1 detector is a Ring Image
Cherenkov detector (RICH), which provides electron/hadron separation
from \pt\ of a few hundred~MeV/$c$ to about 5\,GeV/$c$. The last layer
of the spectrometer is an electromagnetic calorimeter which is used
for photon measurements and electron/hadron separation.

The VTX was commissioned in Run 2011 using \p~collision data at
$\sqrt{s}$\,=\,500\,GeV.  Commissioning was followed by a period of
data collection, which forms the basis of the physics measurements
presented here, in 2011 (for \Au) and 2012 (for \p) RHIC runs.  The
VTX detector consists of four layers of barrel detectors which cover
$\mid$$\eta$$\mid$$<$1.2 and almost 2$\pi$ in azimuth. The inner two
barrels consist of a silicon pixel device with 50$\times$425\,$\mu$m
pixel size. The outer barrels consist of silicon strip detectors with
stereoscopic strips of 80\,$\mu$m~$\times$~3\,cm, these
devices achieve an effective pixel size of
80\,$\times$\,1000\,$\mu$m. 
Details of the VTX can be found in Ref.~\cite{RN2011}.
\begin{figure}[!]                                      
\includegraphics[width=1\textwidth]{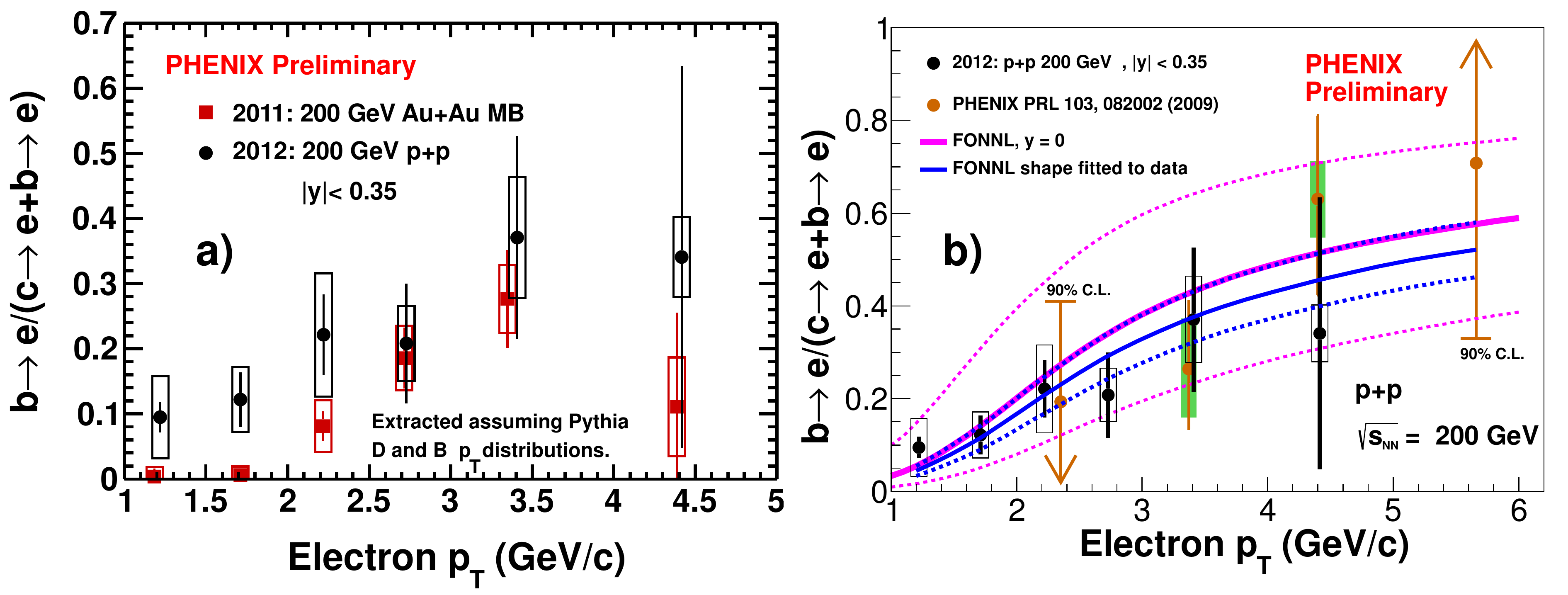}
\vskip -0.15cm
\caption{\small \label{fig2} (a) Fraction of $b$$\rightarrow$$e$ to
  the total ($c$$\rightarrow$$e$ + $b$$\rightarrow$$e$) as a function
  of electron~\pe\ for minimum-bias \Au\ and \p\ at \snn\,200 GeV.
  Panel (a) shows a comparison between the two systems. These
  extracted fractions were obtained assuming {\sc Pythia} $D$ and $B$
  \pt\ distributions. \Au\ data include no uncertainties from modified
  \pt\ spectra.  Panel (b) shows a comparison between \p\ and a FONLL~\cite{Ca2005}.}
\end{figure}                                      
\section{Measurement of the Nuclear Modification Factor for Charm and Bottom}
The first step in measuring $R_{AA}$ for $c$ and $b$ is to determine
the distance of closest approach (DCA) of the candidate electrons to
the primary collision vertex.  As electrons from $D$ and $B$ mesons
decays do not originate at the primary collision vertex, they have a
large DCA.  For example, the decay-lengths of $D^{0}$ and $B^{0}$ are
123~and~457\,$\mu$m respectively. This differentiation using the DCA
is only made possible by the tracking through the VTX.  In the present
measurements, a precise primary collision vertex \textemdash\ ${\rm
  0-5\%}$ central \Au\ collisions at \snn\,200\,GeV has a resolution
($\sigma_x$, $\sigma_y$, $\sigma_z$) of (54$\pm$2, 37$\pm$2,
68$\pm$2)\,$\mu$m \textemdash\ is determined by stand-alone-tracking
using the VTX. These resolutions will be improved by future fine
adjustments to the alignment of the detector.
\begin{figure}[!]                                      
\includegraphics[width=0.85\textwidth]{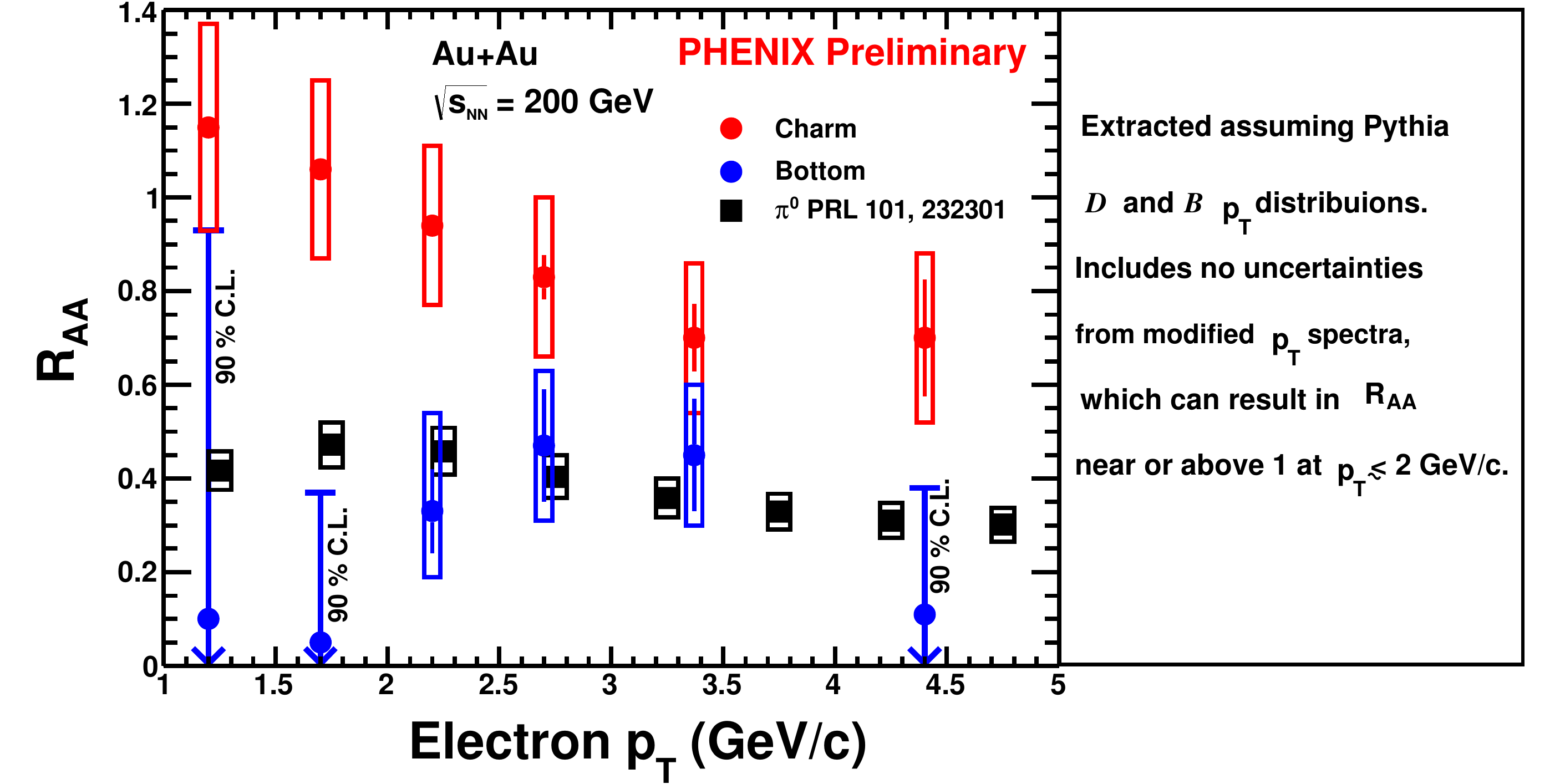}
\vskip -0.3cm
\caption{\small \label{fig3} \Raa\ of electrons from charm and bottom decays
  in \Au\ collisions at \snn\,200\,GeV. The results are compared to
  published PHENIX ${\pi^0}$~\cite{PHEPi0}.}
\end{figure}  

Once candidate electrons are determined from the original PHENIX
central arm components, they are matching to a set of clusters in the
VTX.  The momentum of the track is measured by the Drift
Chamber. Electron identification is done by RICH and energy momentum
matching. The VTX-portion of the reconstructed candidate is used to
determine the DCA to the collision vertex allows the statistical
extraction of flavor identification. Figure~\ref{fig1} shows the raw
DCA distributions of electrons measured at \snn\,200\,GeV for both
minimum bias \Au\ (panel a) and \p\ (b) collisions in the range
2$<$\pe$<$2.5\,GeV/$c$.  The DCA resolutions are estimated to be 70
and 138 $\mu$m, in \Au\ and \p\ respectively.  In the present work,
the DCA resolution for \p\ was determined relative to the beam center,
rather than the collision vertex as used with \Au\ collisions, which
avoids auto-correlations due to the low multiplicity.

The curves presented in Fig.~\ref{fig1} represent the decomposition of
the DCA distributions from all expected signal and background
contributions.  For each \pe\ bin, we first define a DCA fit function,
which has contributions from photonic, charm, bottom, \K, and a
constant-BG, from which the component yields are obtained.  The amount
of hadron contamination was determined using the RICH swap method -- a
standard procedure of PHENIX electron analyses~\cite{bott}. The DCA
distributions of the photonic components (Dalitz, photon conversion)
are Gaussian distributions, which was confirmed with a full {\sc
  Geant} simulation of the PHENIX detector. We additionally studied
the effect of high multiplicities (expected in \Au\ collisions) on the
DCA distribution by embedding simulated electrons and pions in real
\Au\ data. This study indicated that the effect on the DCA due to the
high occupancy is small and justifies the use of a pure Gaussian
distribution as the shape of the photonic component.  The DCA
distribution shape of the $c$, $b$, and \K\ decay components were
determined by simulation.  The $c$ and $b$ distribution shapes were
generated using {\sc Pythia} ($b$$\rightarrow$$e$ and
$c$$\rightarrow$$e$) decays convoluted with the DCA resolution of the
VTX detector. This was cross checked using a full simulation.  The
\K\ background normalization is determined from the number of charged
hadrons in the same \pt\ bin, the measured on $K/h$ ratio, the
branching ratios of \K, and a simulation of \K\ decays.  There are
additional backgrounds at large DCA in the data that are caused by
conversion electrons in the outer VTX layers.  In low multiplicity
events, these outer layer conversions are removed by the requirement
of hits in the first two barrels B0 and B1 of the VTX.  However, in
\Au\ collisions, some of these conversions are confirmed by random
hits in the VTX. Since they are caused by random matching with
unrelated hits in the VTX, their DCA distribution is broad, almost
flat.  This random background was evaluated using a “sideband method”
which showed that the distribution is consistent with flat DCA along
$z$ in the region 0.75 $<$ $\mid$DCA$_Z$$\mid$ $<$1.0\,mm.

The total number of inclusive electrons, $N_{e}^{inc}$, are obtained
after subtracting the backgrounds (hadron contamination, \K, and
constant BG) from the total number electrons in the DCA distribution.
The number of $b$ and $c$ are obtained using $N^{b} = N_{e}^{inc}
\times R_{HQ} \times F^{b}$ and $N^{c} = N_{e}^{inc} \times R_{HQ}
\times (1-F^{b}$), respectively. Here \Fb\ is the fraction of $b$
in heavy flavor electrons,
$F^{b}$=$b\rightarrow$$e$/($c\rightarrow$$e$+$b\rightarrow$$e$),
which is the only unknown parameter. The $R_{HQ}$ is defined as $R_{HQ}
= N_{e}^{HF}/N_{e}^{inc}$ which was estimated by two methods: one based
on 2004 data of photonic/non-photonic ratio and the other based on
measured conversion tagging probability using 2011 data.

Figure~\ref{fig2}a shows the resulting bottom fraction,
$F^{b}$=$b\rightarrow$$e$/($c\rightarrow$$e$+$b\rightarrow$$e$),
as a function of electron \pe\ measured in MB \Au\ and for \p\ 
collisions at \snn\,200\,GeV. The ratio in \p\ collisions is
compared to a fixed-order-plus-next-to-leading-log (FONLL)
perturbative QCD calculation (pink curves)~\cite{Ca2005} as shown in
Fig.~\ref{fig2}b. We observe that FONLL is in good agreement
with current and previous published~\cite{bott} PHENIX results.
Owing to the large uncertainties in the \p\ data the FONLL
calculation shape is varied to fit all experimental data
points of \Fb\ (FONLL blue solid curve).  The blue dashed
curves represent 1$\sigma$ uncertainties from the FONLL shape
fit procedure.  The \Fb\ ratio is used to form $R_{AA}$ for
$c$ and $b$ using Eqn.~\ref{eqn:raa}:
\begin{equation}
\label{eqn:raa}
R^{b\rightarrow e}_{AA} = R^{b+c\rightarrow
  e}_{AA}\frac{F^{b}_{AA}}{F^{b}_{ppFit}}
\quad \quad {\rm and} \quad \quad \quad  
R^{c\rightarrow e}_{AA} = R^{b+c\rightarrow
  e}_{AA}\frac{1-F^{b}_{AA}}{1-F^{b}_{ppFit}}
\end{equation}
where ${R^{b+c\rightarrow e}_{AA}}$ is the nuclear modification factor
for single electrons heavy flavor measured by PHENIX~\cite{PPG77}.  We
observe that the nuclear modification of $c$ is less than that for
$\pi^{0}$s ($R^{c\rightarrow e}_{AA}$$>$$R^{\pi^{0}}_{AA}$), as shown
in Fig.~\ref{fig3}.

The bottom/charm separation results, shown in Figs~\ref{fig2} and
\ref{fig3}, are extracted assuming {\sc Pythia} $D$ and $B$
\pt\ distributions. These results demonstrate that the \Au\ data are
inconsistent with these input assumptions unless there is also a large
suppression of electrons from bottom across the measured \pe\ range.
This large suppression implies a large change in the parent hadron
\pt\ distributions, which results in changes in the electron DCA
distributions. This causes an additional set of uncertainties in \Rb\
which are not included in Figure~\ref{fig3}. Because the charm fraction
dominates at low \pe, the \Rc\ is less affected.  We are actively
working on evaluation of these uncertainties.  These results imply
that either a large suppression of $ b \rightarrow e$ or a large
modification of $B$ meson \pt\ distributions, which implies very
interesting physics of $B$ mesons in \Au\ collisions.
\section{Summary}
The DCA distributions of single electrons have been measured by the
VTX detector in \p\ and \Au\ collisions at \snn\ 200 GeV. By selecting
electrons with a DCA to the primary vertex larger than 100 $\mu$m,
the photonic electron background was suppressed by orders of
magnitude. This suppression results in a clean and robust measurement
of heavy flavor production in the single electron channel. From this,
the relative fraction of $b$ and $c$ quarks is determined. In \p, the
FONLL perturbative QCD calculation of $b\rightarrow e /(c\rightarrow
e+b\rightarrow e$) ratio is in good agreement with the data. This
allows us to measure $R_{AA}$ for single electrons from $D$ and $B$
decays. We observe a higher $R_{AA}$, i.e. less suppression, for $D$
mesons compared that observed for $\pi^{0}$s.  While for $B$, the data
imply a large suppression of $b \rightarrow e$ or a large modification
of $B$ meson \pt\ distributions. It should be noted that the
bottom/charm separation results are extracted assuming {\sc Pythia}
$D$ and $B$ \pt\ distribution. This leads to additional uncertainties
in \Raa which are not included in the present results; PHENIX
Collaboration is actively working on evaluation of the uncertainties.

\end{document}